# CNSMO: A Network Services Manager/Orchestrator Tool for Cloud Federated Environments

J. Aznar, E. Escalona, I. Canyameres, O. Moya, A. Viñes

Fundació I2CAT, Barcelona 08034, Spain

E-mail: {jose.aznar, eduard.escalona, isart.canyameres, oscar.moya, Albert.vinyes}@i2cat.net

*Abstract*— Application service providers (ASPs) now develop, deploy, and maintain complex computing platforms within multiple cloud infrastructures to improve resilience, responsiveness and elasticity of their applications. On the other hand, complex applications have little control and visibility over network resources, and need to use low-level hacks to extract network properties and prioritize traffic. This biased view, limits tenants´ flexibility while deploying their applications and prevents them from implementing part of the application logic in the network. In this paper, we propose the CNSMO (CYCLONE Network Services Manager/Orchestrator) tool to bring the innovation at federated cloud environments by bridging these network service capabilities to cloud based services as part of the overall CYCLONE solution. The integration of networking aspects with purely federated clouds, will allow users to request specific infrastructures and manage their dedicated set of coordinated network and IT resources in an easy and transparent way while operating dynamic deployments of distributed applications.

## I. Making the most of the network in Cloud federated environments

Future Internet (FI) frameworks and architectures are adopting service-oriented approaches in which IT and Network coordination is a must. More specifically, in Cloud federated scenarios, a group of Cloud infrastructure providers are federated and trade their surplus resources amongst each other to gain economies of scale, efficient use of their assets, and expansion of their capabilities, e.g. to overcome resource limitations (scalability), avoid vendor lock-in, ensure availability and recovery, provide with geographic distribution to ensure low latency, regulatory constraints, cost efficiency and energy saving, etc. [1]. Thus, this model enables to deliver computing facilities to Service Providers (SPs) using resources of either one infrastructure provider, or combination of different providers.

However, the different stakeholders involved in the value chain also experience a number of barriers that should be lowered down to ensure a proper adoption: This is the case of both academic and commercial Application Service Providers (ASPs) which have embraced Infrastructure-as-a-Service (IaaS) cloud infrastructures because of their elasticity, flexibility, and dynamic resource provisioning. These application service providers have moved beyond simple, single-machine applications and now develop complex computing platforms that are deployed and maintained within the cloud. These complex applications are often distributed between cloud infrastructures to provide resilience against cloud provider outages, higher levels of elasticity, and better response times for their users by placing services near the clients. Moreover, they are often designed to scale automatically in response to demand and to permit live upgrades of the underlying software. Nevertheless, ASPs have barely been in control of the network resources, limiting tenants' flexibility while deploying such applications. Also Network Service Providers (NSPs) become limited, since they experience little control or complete lack of knowledge on the semantics of the application data and on the network requirements that the applications may request as part of the application deployment. Additionally, the variation of network configuration capabilities among cloud platforms entails significant constraints while provisioning inter-cloud network federated resources. Previous limitations highlight the important role that the network plays in cloud federated scenarios.

Thus, it is clear that such Cloud federated scenarios impose challenging requirements and essential characteristics that fall on the network aspects of service delivery in terms of service delivery automation, resource management or on-demand creation of network connectivity between Communication Service Providers (CSPs) and Communication Service Consumers (CSCs), within Data Centers (DCs) or between the computing nodes of a CSPs infrastructure. All these capabilities to offer network connectivity as a Service is what in the literature refers to "Network-as-a-Service" (NaaS). NaaS is a business and service delivery model related to network infrastructure for providing network services with dynamic and scalable, yet secure and isolated, access to networks for multiple tenants.

The H2020 CYCLONE project [2], is currently addressing ASPs networking requirements to ease and automate applications deployment in Cloud Federated infrastructures, bringing innovation by bridging the network service capabilities to cloud based services, extending the IaaS model and integrating the networking aspects and requirements of cloud federations in order to provide users with the means to include also network management options in the services' requests. The possibility to specify network requirements while deploying complex applications opens a wide range of possibilities for ASPs. This paper, explores the cloud complex applications requirements in terms of cloud networking

(section II) and proposes the Cyclone Network Services Manager/ Orchestrator (CNSMO), a refactored version of the OpenNaaS software [3] to automatically integrate and deploy network services in cloud federations (Section III). Moreover, it is explained the coexistence and integration with other critical aspects while serving cloud applications, especially in terms of cloud service platforms and proposes a generic service model (section IV). Most relevant finding and ongoing work is summarized in section V.

## II. CLOUD COMPLEX APPLICATION REQUIREMENTS IN TERMS OF CLOUD NETWORKING

In order to retrieve functional networking requirements imposed by complex applications in cloud federation environments, CYCLONE is considering two flagship domains: academic use cases for bioinformatics research and use cases for a commercial deployment for smart grids in the energy sector. These are guiding the gathering of the requirements and initial development of the network services in the CNSMO platform.

### A. Use cases overview

The Bioinformatics use case deals with the collection and efficient analysis of biological data, particularly genomic information from DNA sequencers. Bioinformatics software is characterized by a high degree of fragmentation: literally hundreds of different software packages are regularly used for scientific analyses with an incompatible variety of dependencies and a broad range of resource requirements. For this reason, the bioinformatics community has strongly embraced cloud computing with its ability to provide customized execution environments and dynamic resource allocation.

The energy sector use case deals with the "202020" climate change mitigation goals of the European Union [4]. The huge amount of decentralized energy components included into the grid forces the grid to become smarter. Thus, integrating the latest Information and Communication Technology (ICT) for managing the energy components in the grid is becoming more and more essential. ICT from distributed embedded control to Big Data and Cloud Computing is essential for the transition. Additional information and a full description of the use cases are provided in [5].

### B. Complex applications' network requirements

In this section we provide the most relevant requirements that previous complex applications demand in cloud federated scenarios in terms of networking, taking as reference the CYCLONE project user cases. Table 1 shows them:

| ID | Network requirement | Bioinformatics | 202020 Energy Grid |
|---|---|---|---|
| 1 | End-to-end secure data management | ✓ | ✓ |
| 2 | VM isolated network | ✓ | |
| 3 | VPN connectivity services | ✓ | |
| 4 | Multi-cloud distribution of community reference datasets | ✓ | |
| 5 | Dynamic network resource allocation | ✓ | ✓ |
| 6 | Multi-clouds distribution of user data | ✓ | ✓ |
| 7 | WAN high bandwidth links | ✓ | ✓ |
| 8 | Guaranteed network performances (QoS) | ✓ | ✓ |

Table 1. Summary of most important cloud networking requirements retrieved from the CYCLONE use cases.

*1) Dynamic network resource allocation*

For enabling the aggregation, calculation and visualization of applications to the distributed multi-cloud environment, different network resources need to be allocated dynamically according to the requirements of the different steps of the application workflow.

*2) Multi-tenancy support*

The possibility to enable different types of tenants over the same network resources is achieved by means of network virtualization and network isolation. Enabling different application deployments utilizing common network resources constitutes an added-value feature to IaaS providers since it enables a better utilization of system resources. This feature applies not only to intra-DC domains but it is also extensible to inter-DC networks.

*3) VM isolated network*

The VMs of a user must be deployed in a dedicated and isolated network that could be reached only by its owner, according to its identity and access credentials. This will isolate its own resources (VMs and data) from other users, and also avoid any impact with other external VMs.

*4) VPN connectivity services*

A life science researcher needs to access all its VMs in a simple manner through a single point of entry such as VPN services. Such features should be presented to the user ideally through a web service.

*5) SDN/OpenFlow support*

With the recent adoption of Software Defined Networking (SDN) technologies within Data Centers, it is also desirable to include the possibility to configure and manage networking resources which operate based on SDN/OpenFlow (OF).

*6) Inter DC connectivity*

Although previously mentioned in some other requirements, inter-DC connectivity requires a separate subsection. Federated Cloud services may need to rely on dedicated inter-DC connectivity services that can either based on connection-oriented paradigms and QoS in controlled and deterministic ways (e.g., through IP/VPN, MPLS, WAN, dedicated leased lines, or optical circuits).

*7) Multi-clouds distribution of community reference datasets*

The deployment of bioinformatics workflows over two or more cloud infrastructures requires that the collections of public reference data used during the treatment are available in all of these clouds. These public datasets are, for example, the European Nucleotide Archive, the Ensemble Genomes resource, the Human Genome version 19 hg19 [hg19] or

GRCh37 [GRCh37] and the UniProt (Universal Protein Resource) catalogue of information on proteins. These datasets are public data that do not need to have security rules associated to their access or transfer. However, they require automatic replication mechanisms to be deployed on several cloud infrastructures and to be updated each time the reference dataset is updated on the bioinformatics site. In terms of inter-DCs networks, that means having a high performance network to replicate hundredths of gigabytes and to pass the firewalls without performance loss between the different sites. In terms of local networks, a high-performance and distributed file system should be mounted in all the cloud users VMs.

*8) Guaranteed network performance (QoS)*

Some bioinformatics tools used to analyze genomics data require graphical display with X11 technology, for example the software IGV – Integrative Genomics Viewer. Associated QoS should be enabled to the link between the user LAN and the DC, to satisfy the X11 remote display requirements in terms stability, priority and performances. In the energy domain, a connection that satisfies the QoS parameters in terms of bandwidth and latency is required for the communication with the SCI-platform. Guaranteeing the QoS (mainly BW and latency) is a requirement coming from the applications' specifications that entail traffic engineering mechanisms at the network level.

## III. OPENNAAS/CNSMO SYSTEM ARCHITECTURE FOR NETWORK SERVICES

Previous stated requirements have been mapped into concrete network services to be integrated as part of the overall CYCLONE platform solution to ease ASPs applications deployment, configuration and management.

CNSMO is the software component responsible of deploying, configuring and running the networking services. CNSMO is a lightweight micro-services framework developed at i2CAT. Leveraging the base concepts of Apache Mesos, CNSMO is a lightweight distributed platform defining a basic service API and service life-cycle, together with an inter-service communication mechanism implementing the actor model and supporting different communication protocols. The system is capable of deploying and running multiple services in both local and remote environments.

Although services may run each in a dedicated VM or container, due to requirements imposed by the Cyclone project, Cyclone networking services run directly on the user-space of user defined VMs. These services have been carefully designed to have a small fingerprint in the VMs they run, with the goal to minimize the impact on the user space. This is achieved by means of docker containers.

As of now, the only available service is a multi-cloud VPN service orchestrating the deployment of a VPN in a pre-selected set of VMs running in different clouds, which allows secure inter-cloud communication. A firewalling service and a load balancer are currently under development.

### A. CNSMO architecture and software components

This section describes the architectural modules that compose the CNSMO application, their rationale and how they interact with each other. Basically, CNSMO is based in micro-services architecture, with the following components:

- The **Northbound API** exposes a set of methods for management of the overall module, but also for the management of the services implemented by specific network services modules that are present. It acts as the CNSMO single management entry-point. Thus, this is the only module other elements of the Cyclone ecosystem interact with.

- The **Core agent** is the glue of the system and the engine that manages the micro-services ensuring all pieces of the system work together as expected. It tracks network services modules in the system, uses distributed state storage and in general acts as a bridge between the rest of components.

- The **Message Queue (MQ)** is the messaging system used by components to communicate to each other. It offers serialization mechanisms providing implementation independency between modules. It also offers load balancing features, useful to distribute the application load between multiple instances of the same CNSMO component, in this case, multiple instances of the same specific network service module (e.g., VPN service module).

- The **specific network service modules** (one per network service) are the ones providing networking features to the end-user. Each module offers a specific networking service (e.g., VPN, firewall, load balancer, etc.) and takes the responsibility of deploying such service where it is said to. Information related to design and implementation of specific service modules is out of the scope of this paper.

- **Distributed state storage** is used to store the application state, so it can recover from failure or system shutdown. By using it intensively, module can become stateless which turns out to be very useful when scaling.

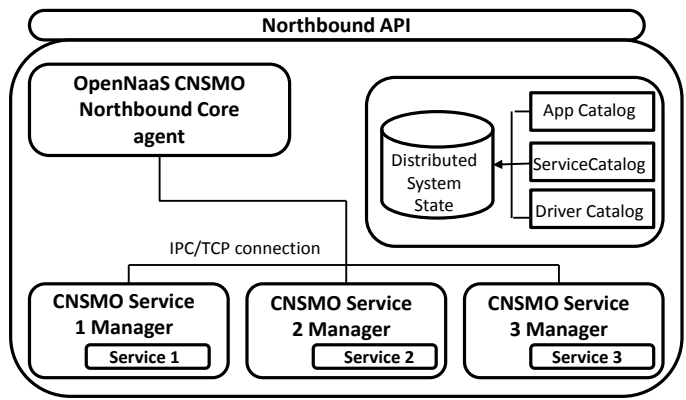

Figure 1. CNSMO framework platform.

## IV. NETWORK SERVICES DEPLOYMENT IN CLOUD FEDERATED SCENARIOS

### A. OpenNaaS/CNSMO service concept

The CNSMO services concept has been designed so that they are recursive, stateless, independent and developed under the premise that a service is potentially to launch any other service present developed in CNSMO. Figure 2 shows the shape of a CNSMO core service scheme. Additionally, in the right side of the same figure it can be shown an example for a concrete network service (VPN). The structure of the service is as follows:

- **Service API**: Exposes a set of methods to remotely manage the service lifecycle.
- **System state API**: Exposes a set of methods to remotely store and react to changes in the state of each system component.
- **Deployment API**: Exposes a set of methods to remotely deploy the service.
- **System.d/CLI driver**: Driver telling the framework how to launch/deploy the service.
- **Service management logic**: Internal service logic. It may stablish a VPN or manage firewalling rules among others, depending on the service intention.

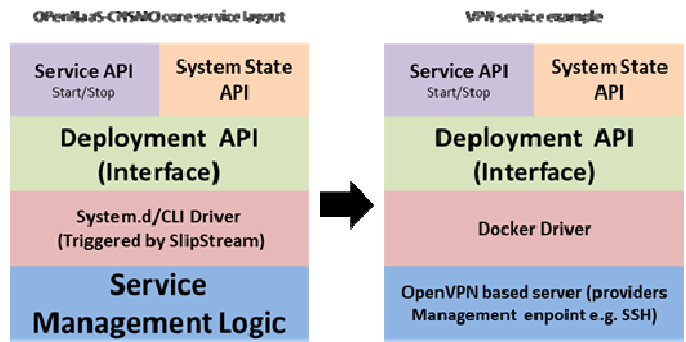

Figure 2. CNSMO core service layout and VPN service example mapping.

### B. CNSMO Impact onother application components

Applications willing to take advantage of CNSMO networking services, must explicitly add the CNSMO component to the application. Specifically, it is required to have a CNSMO agent installed and it must have the appropriate deployment scripts to trigger the CNSMO agent so it gets coordinated with CNSMO components. In order to ease this task, the CNSMO relies on another CYCLONE tool: SlipStream [6]. SlipStream is a multi-cloud application management platform used by CYCLONE to handle the automated deployment and full application management lifecycle, including the deployment, testing, certification and optimization, within Infrastructure as a Service (IaaS) cloud providers. SlipStream provides a general cloud application model and an associated abstract API to hide differences between cloud service providers from the cloud application developers and operators, making cloud portability and multi-cloud applications a reality. Also, it provides connectors that enable to support a large number of Infrastructure as a Service (IaaS) technologies and services (e.g. AWS, OpenStack, StratusLab, Azure, etc.).

Thus, in the context of the CYCLONE project, the CNSMO is an application component in SlipStream jargon available in its catalog. CNSMO enabled SlipStream components (CNSMO-agent) will be created, containing a base distro with the CNSMO agent and deployment scripts. CNSMO-agent images may be created from multiple distros if required. These components are meant to be used as parent images in other components. It is important that deployment scripts defined in CNSMO-agent components run before the deployment scripts of children components. This is so, because networking services may enable networking interfaces that children components must use. Application components willing to make use of CNSMO-agent component, must define the CNSMO-agent as its parent image. This requires that any further software installation must be performed through SlipStream installation and deployment scripts.

It has also been considered the possibility in which this approach is not valid from the ASP perspective. In such a case, the application developer would need to pre-install the CNSMO-agent into the component image and manually add the CNSMO-agent deployment scripts to its component scripts ensuring that they are executed in the correct order. That is, CNSMO-agent deployment scripts are executed before any other configuration script of the component.

### C. Bootstrapping process: Running the network services.

Starting from the premise that CNSMO relies on SlipStream to deploy the network services in cloud federated environments, the bootstrapping process to run a network service is shown in Figure 3. As a reference, the VPN network service has been taken. In a first step, the application developer needs to specify the application profile and requirements. To this end, the user must fill in a recipe in which the details of the required deployment (including network services) is specified. SlipStream deploys the requested VMs to run user's application based on the requested recipe (step 2 of Figure 3).

CNSMO bootstrap is triggered by SlipStream and its deploying recipe. Each time an application is deployed by SlipStream, it builds component images, runs those images in VMs and, afterwards, runs deployment recipes of each VM. The CNSMO image contains a *system.d* service that runs CNSMO when the system loads. When launched, CNSMO offers an API so that others can interact with it. The CNSMO SlipStream application contains a deployment recipe with appropriate instructions for CSNMO to deploy the chosen networking services. The recipe is run by SlipStream inside the CNSMO VM. This recipe gathers information of the deployment from SlipStream by means of Service State client, calls CNSMO API to deploy desired network services and uses the service state client to announce to the rest of component that the networking services are set up, so

SlipStream can resume their deployment (running their deployment script).

The deployment information is used by CNSMO to determine which services must be deployed in each component. CNSMO locates the rest of the components in the deployment, finds their roles, determines which service must be deployed in each component and deploys them (Step 4). Figure 3 shows the high level deployment workflow for a VPN service following the described bootstrapping sequence.

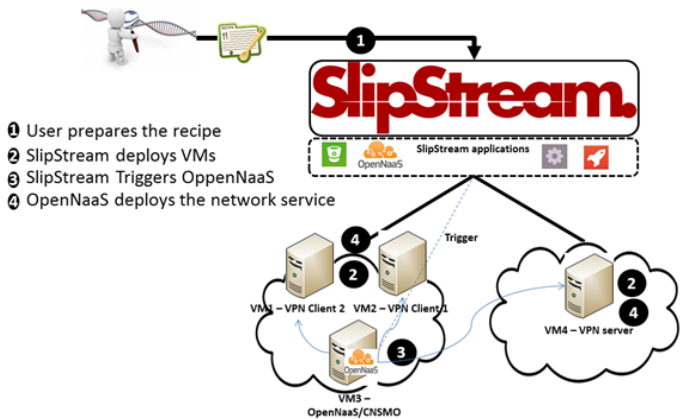

Figure 3. VPN service deployment. 2 openVPN clients and 1 openVPN server are installed in the application VMs as docker containers.

## V. NETWORK SERVICES DEMONSTRATION

Previous sections have explained the CNSMO architecture, the approach to design network services and how to bootstrap and deploy them in the context of the CYCLONE project, relying on SlipStream. Currently, it has proven the integration of the VPN service as part of the network services that can be deployed by utilizing SlipStream. The experienced delay while deploying the VPN service together with the application is negligible comparing to the deployment of an application which does not include the VPN service. Thus, there is no impact on the application provisioning time, which is highly appreciated by the users (ASPs). Figure 4 shows the web panel while launching a network service developed with OpenNaaS/CNSMO tool, integrated with the SlipStream Software as a Service (SaaS) online tool, Nuvla [7]. As it has been previously pointed, Slipstream has connectors to a large variety of Cloud Service Platforms (CSPs). For the demonstration of the network services, OpenStack and StratusLab CSPs private IaaS have been utilized.

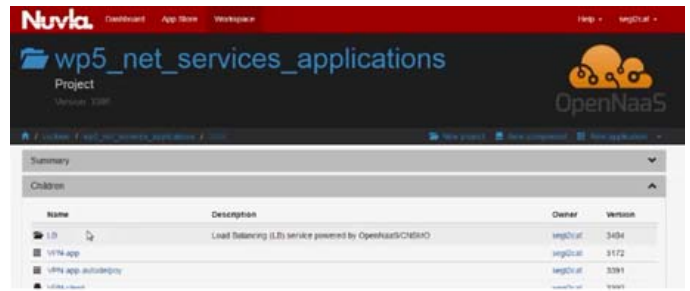

Figure 4. SlipStream online SaaS Nuvla front end integrated with OpenNaaS/CNSMO tool.

## VI. FUTURE WORK

OpenNaaS CNSMO has been conceived as a network services implementation and deployment tool for cloud federated and distributed environments, with different CSP flavors and using SlipStream as application deployment and management software tool. A VPN service that addresses requirements #2 and #3 of the use cases requirements table (see Table 1) has been implemented in order to validate the CNSMO architecture, the integration with SlipStream and the deployment of the network services together with the application, without significantly impacting the process.

Future immediate work includes the implementation of other network services (firewalling and load balancing mechanisms) and other services to address additional requirements retrieved from the current and future use cases. The automation of certain integrated mechanisms between SlipStream and CNSMO is also an important requirement in the path towards "*zero touch provisioning*". Finally, the integration with concrete security mechanisms and requirements is also part of the ongoing work.

ACKNOWLEDGMENT

The work presented in this paper is supported by the EU funded Horizon2020 project CYCLONE under Grant agreement no. 644925